\documentstyle[sprocl]{article}

\font\eightrm=cmr8

\def\Journal#1#2#3#4{{#1} {\bf #2}, #3 (#4)}

\def\NCA{\em Nuovo Cimento}
\def\NIM{\em Nucl. Instrum. Methods}
\def\NIMA{{\em Nucl. Instrum. Methods} A}
\def\NPB{{\em Nucl. Phys.} B}
\def\NPA{{\em Nucl. Phys.} A}
\def\PLB{{\em Phys. Lett.}  B}
\def\PRL{\em Phys. Rev. Lett.}
\def\PRD{{\em Phys. Rev.} D}
\def\ZPC{{\em Z. Phys.} C}
\def\PRP{\em Phys. Rep.}
\def\EPJC{{\em Eur. Phys. J.} C}

\def\st{\scriptstyle}
\def\sst{\scriptscriptstyle}
\def\mco{\multicolumn}
\def\epp{\epsilon^{\prime}}
\def\vep{\varepsilon}
\def\ra{\rightarrow}
\def\ppg{\pi^+\pi^-\gamma}
\def\vp{{\bf p}}
\def\ko{K^0}
\def\kb{\bar{K^0}}
\def\al{\alpha}
\def\ab{\bar{\alpha}}
\def\be{\begin{equation}}
\def\ee{\end{equation}}
\def\bea{\begin{eqnarray}}
\def\eea{\end{eqnarray}}
\def\CPbar{\hbox{{\rm CP}\hskip-1.80em{/}}}

\begin{document}

\title{REAL-TIME SIMULATIONS OF HIGH-ENERGY NUCLEAR COLLISIONS}

\author{ALEX KRASNITZ}

\address{UCEH, Universidade do Algarve, P-8000 Faro, Portugal}

\author{RAJU VENUGOPALAN}

\address{Niels Bohr Institute, Blegdamsvej 17, Copenhagen, Denmark, DK-2100}


\maketitle\abstracts{We discuss real time simulations of high energy nuclear 
collisions in a classical effective theory of QCD at small x. At high 
transverse momenta, our results match the lowest order predictions of 
pQCD based mini--jet calculations. We discuss novel non--perturbative 
behaviour of the small x
modes seen at small transverse momenta.}

\section{Introduction}

The space--time evolution of nuclei in high energy heavy ion 
collisions and the 
various proposed signatures of the hot and 
dense matter formed~\cite{QM96} 
depend sensitively on the initial conditions for the evolution~\cite{comment}.
These are the distributions of partons in each of 
the nuclei {\it prior} to the 
collision. In the standard perturbative QCD approach, 
observables from the collision are computed by convolving
the known parton distributions of each nucleus  
with the elementary parton--parton 
scattering cross sections. At the RHIC and LHC 
colliders, hundreds of 
mini--jets may be formed in the initial 
collision~\cite{KajLanLin}. 
Final state interactions of these mini--jets are often
described in multiple scattering models (see 
Ref.~\cite{Wang} and references therein) or in classical cascade 
approaches (see Ref.~\cite{Geiger} and references therein).
The possible ``quenching'' of these mini--jets has also been studied 
and proposed as a signature of the formation of a quark gluon plasma~\cite
{GyuPluWang}. 
Recently, initial conditions for the energy density and velocity obtained 
in the mini--jet approach have been used in a simple hydrodynamic model to 
study the late time evolution of matter in high energy nuclear
collisions~\cite{KKV}.

In this paper, we describe an {\it ab initio} QCD based approach to the
study of high energy nuclear collisions. This model 
naturally incorporates coherence effects which
are important at small x and small transverse momenta and
simultaneously reproduces 
the standard mini--jet results at large transverse momenta. It also
contains a self--consistent space--time
picture of the nuclear collison. The model is based on an effective action
approach to QCD initially developed by McLerran and 
Venugopalan~\cite{RajLar},  
and later further developed by Ayala, Jalilian--Marian, McLerran and 
Venugopalan~\cite{AJMV} and by 
Jalilian--Marian, Kovner, Leonidov and Weigert~\cite{JKMW,JKLW,JKW}.

The above mentioned effective action contains one dimensionful parameter,
$\mu$,
the total color charge squared per unit area. 
The classical fields corresponding to the saddle point solutions 
of the effective theory are the non--Abelian
analogue of the Weizs\"acker--Williams fields in 
electrodynamics. Exact analytical expressions for these fields have been
obtained recently~\cite{JKMW,Kovchegov}. Further, it has been shown
explicitly that $\mu$ obeys renormalization group equations in rapidity 
$y$ and momentum transfer squared $Q^2$.
These reduce to the well known BFKL and DGLAP equations
respectively in the appropriate limits~\cite{JKLW,JKW}.

The above model was applied to the problem of nuclear collisions by
Kovner, McLerran and Weigert. They formulated the problem as the collision of
Weizs\"acker--Williams fields~\cite{KLW}. 
Further, classical gluon radiation corresonding to perturbative
modes was studied by these authors and later in greater detail by several
others~\cite{gyulassy,DirkYuri,SerBerDir}. In the small x limit, 
the classical Yang--Mills result agrees with the quantum Bremsstrahlung 
result of Gunion and Bertsch~\cite{GunionBertsch}.

The perturbative approach is of course very relevant and useful. However, it
is essential to consider the full non--perturbative approach for the
following reasons. Firstly, the classical gluon radiation computed
perturbatively is infrared singular, as is also the case in mini-jet
calculations, and therefore rather sensitive to the
cut--off. It was argued in Ref.~\cite{KLW,DirkYuri} that a
natural scale for the cut-off is $k_t\sim \alpha_S \mu$. However, to
be sure, it is important to perform a full calculation. Secondly, the
non-perturbative approach is crucial to study the space-time evolution of
the nuclei and in particular, the possible thermalization of the system.
This has several ramifications for various quark gluon plasma signatures. For
instance, if thermalization does occur, then hydrodynamic evolution of the
system is reasonable~\cite{Bj}.  In that event, our approach would provide
the necessary initial temperature and velocity profiles for such an
evolution~\cite{KKV}.

We discuss in this paper first results from real time simulations of the full,
non--perturbative evolution of classical non--Abelian Weizs\"acker--Williams
fields. Such a simulation is possible since the fields are classical. Similar
classical simulations have been
performed in the context of sphaleron induced baryon number
violation~\cite{Krasnitz} and chirality violating transitions in hot gauge
theories ~\cite{Moore}. In brief, the idea is as follows~\cite{RajKrasnitz}.
We write down the
lattice Hamiltonian which describes the evolution of the small x classical 
gauge fields.  It is the Kogut--Susskind Hamiltonian in
2+1--dimensions coupled to an adjoint scalar field. The lattice equations of
motion for the fields are determined straightforwardly. The initial
conditions for the evolution are the Weizs\"acker--Williams
fields for the nuclei before the collision. Interestingly, the dependence on
the sources does not enter through the Hamiltonian but
instead through the initial conditions. 

A related approach is that of Mueller, Kovchegov and 
Wallon~\cite{MuellKovWall}, where they combined Mueller's dipole 
picture of high energy scattering~\cite{Mueller} with the 
classical Yang--Mills picture~\cite{Kovchegov} to study nucleon--nucleus 
scattering.  
For alternative approaches, we refer the reader to the work of Makhlin and 
Surdutovich~\cite{MakhSurd}and of Balitsky~\cite{Ian}.

Our paper is organized as follows. In the following section we briefly
discuss the problem of initial conditions for nuclear collisions and the
perturbative computation of classical gluon production.  We also discuss a
non-perturbative Hamiltonian approach to the solution of the full
Yang-Mills equations. In section 3, we formulate the problem of solving the
Yang-Mills equations on the lattice. Starting from the lattice action and
assuming boost invariance, we write down the lattice Hamiltonian, the
Hamilton equations of motion and the initial conditions for the dynamical
fields and their conjugate momenta on the lattice. The initial conditions depend
on a single dimensional parameter $\mu$, the color charge per unit area per unit
rapidity. Numerical results from
our simulations and comparisons to lattice perturbation theory (LPTh) are
discussed in section 4. These are performed for a range of values of $\mu$ =
0.015--0.2, and for lattice sizes from $10\times 10$ to $160\times 160$
measured in units of the lattice spacing. We summarize our results in
section 5.

\section{The Weizs\"acker--Williams approach to high energy nuclear collisions}
\vspace*{0.2cm}

The model of McLerran and Venugopalan of gluon fields in a nucleus at 
small $x$ was applied to nuclear collisions by 
Kovner, McLerran and Weigert~\cite{KLW}. 
We shall now discuss their formulation of the problem and their
perturbative computation, to second order in the parameter $\alpha_S\mu/k_t$, 
of classical gluon radiation in nuclear collisions.
We will then briefly discuss a non--perturbative Hamiltonian approach which
suggests how all orders in $\alpha_S\mu/k_t$ can be computed numerically. 
The implementation of this approach on the lattice is described in 
section 3.

In high energy nuclear collisions, hard valence parton modes
are highly Lorentz contracted, static sources of color charge for the
wee Weizs\"acker--Williams modes in the nuclei. The sources are
described by the current
\be
J^{\nu,a}(r_t) = \delta^{\nu +}g\rho_{+}^a (r_t)\delta(x^-) + \delta^{\nu -}
g\rho_{-}^a (r_t) \delta(x^+) \, ,
\label{sources}
\ee
where $\rho_{+,-}$ are the color charge densities of the hard
modes.  The classical field of the two nuclei is obtained by solving
$D_\mu F^{\mu\nu} = J^\nu$
for the two above mentioned light cone sources.

Gluon distributions are simply related to the Fourier transform $A_i^a (k_t)$ 
of the classical field by $\langle A_i^a(k_t) A_i^a(k_t)\rangle_\rho$. 
The subscript $\rho$ above refers to the averaging over the color charge 
distributions (performed independently 
for each nucleus) with the Gaussian weight $\mu^2$. We assume equal A for
simplicity.

We have omitted the rapidity dependence
of the charge distributions in the equations immediately above. 
The omission is discussed further in the next section. We note here
that the rapidity dependence of the charge distribution is also absent in 
Ref.~\cite{KLW} (see the discussion below Eq.~\ref{yangmill2}).

Before the nuclei collide ($t<0$), a solution of the equations of motion is
$A^{\pm}=0$, 
$A^i= \theta(x^-)\theta(-x^+)\alpha_1^i(r_t)+\theta(x^+)\theta(-x^-)
\alpha_2(r_t)$ where,~\cite{gyulassy}
\be
\alpha_{1,2}^i(r_t) = {1\over {ig}}\left(Pe^{-ig\int_{\pm \eta_{proj}}^0 
d\eta^\prime
{1\over {\nabla_{\perp}^2}}\rho_{\pm}(\eta^\prime,r_t)}\right)^{\dagger} 
\nabla^i\left(Pe^{-ig\int_{\pm \eta_{proj}}^0 d\eta^\prime {1\over 
{\nabla_{\perp}^2}}\rho_{\pm}(\eta^\prime,r_t)}\right) \, .
\label{puresoln}
\ee
Above, $\eta=\eta_{proj}-\log(x^-/x_{proj}^-)$ is the rapidity of the nucleus 
moving along the positive light cone with the gluon field $\alpha_1^i$ and
$\eta=-\eta_{proj}+\log(x_{proj}^+/x^+)$ is the rapidity of the nucleus moving
along the negative light cone with the gluon field $\alpha_2^i$. At 
central rapidities, (or $x\ll 1$) the source density varies
slowly as a function of rapidity and $\alpha^i\equiv \alpha^i(r_t)$.
The above expression suggests that for $t<0$ the solution is simply the
sum of two disconnected pure gauges. Just as in the Weizs\"acker--Williams 
limit in QED, the transverse components of the electric field are 
highly singular. 

For $t>0$ the solution is no longer pure gauge. Working in the Schwinger 
gauge $x^+ A^- + x^- A^+ =0$, 
the  authors of Ref.~\cite{KLW} found that with the ansatz
$A^{\pm}=\pm x^{\pm}\alpha(\tau,r_t)$, 
$A^i=\alpha_\perp^i(\tau,r_t)$,
(where $\tau=\sqrt{2x^+ x^-}$), the Yang--Mills equations could be written in
the simpler form
\bea
{1\over \tau^3}\partial_\tau \tau^3 \partial_\tau \alpha + [D_i,\left[D^i,
\alpha\right]]
&=&0 \, , \nonumber \\
{1\over \tau}[D_i,\partial_\tau \alpha_\perp^i] + ig\tau[\alpha,\partial_
\tau \alpha] &=&0\, ,\nonumber \\
{1\over \tau}\partial_\tau \tau\partial_\tau \alpha_\perp^i
-ig\tau^2[\alpha,\left[D^i,\alpha\right]]-[D^j,F^{ji}]&=&0 \, . 
\label{yangmill2}
\eea 
Note that the above equations of motion are independent of $\eta$--the 
gauge fields in the forward light cone are therefore only functions of
$\tau$ and $r_t$ and are explicitly boost invariant.

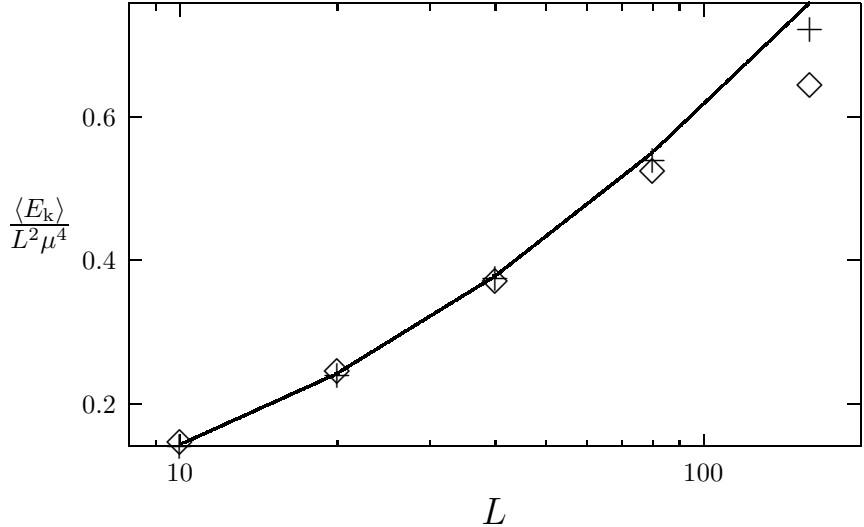
\begin{figure}[ht]
\setlength{\unitlength}{0.240900pt}
\ifx\plotpoint\undefined\newsavebox{\plotpoint}\fi
\sbox{\plotpoint}{\rule[-0.200pt]{0.400pt}{0.400pt}}%
\begin{picture}(1350,900)(0,0)
\begin{Large}
\font\gnuplot=cmr10 at 10pt
\gnuplot
\sbox{\plotpoint}{\rule[-0.200pt]{0.400pt}{0.400pt}}%
\put(181.0,229.0){\rule[-0.200pt]{4.818pt}{0.400pt}}
\put(161,229){\makebox(0,0)[r]{0.2}}
\put(1310.0,229.0){\rule[-0.200pt]{4.818pt}{0.400pt}}
\put(181.0,455.0){\rule[-0.200pt]{4.818pt}{0.400pt}}
\put(161,455){\makebox(0,0)[r]{0.4}}
\put(1310.0,455.0){\rule[-0.200pt]{4.818pt}{0.400pt}}
\put(181.0,680.0){\rule[-0.200pt]{4.818pt}{0.400pt}}
\put(161,680){\makebox(0,0)[r]{0.6}}
\put(1310.0,680.0){\rule[-0.200pt]{4.818pt}{0.400pt}}
\put(181.0,163.0){\rule[-0.200pt]{0.400pt}{2.409pt}}
\put(181.0,849.0){\rule[-0.200pt]{0.400pt}{2.409pt}}
\put(223.0,163.0){\rule[-0.200pt]{0.400pt}{2.409pt}}
\put(223.0,849.0){\rule[-0.200pt]{0.400pt}{2.409pt}}
\put(261.0,163.0){\rule[-0.200pt]{0.400pt}{4.818pt}}
\put(261,122){\makebox(0,0){10}}
\put(261.0,839.0){\rule[-0.200pt]{0.400pt}{4.818pt}}
\put(508.0,163.0){\rule[-0.200pt]{0.400pt}{2.409pt}}
\put(508.0,849.0){\rule[-0.200pt]{0.400pt}{2.409pt}}
\put(653.0,163.0){\rule[-0.200pt]{0.400pt}{2.409pt}}
\put(653.0,849.0){\rule[-0.200pt]{0.400pt}{2.409pt}}
\put(756.0,163.0){\rule[-0.200pt]{0.400pt}{2.409pt}}
\put(756.0,849.0){\rule[-0.200pt]{0.400pt}{2.409pt}}
\put(835.0,163.0){\rule[-0.200pt]{0.400pt}{2.409pt}}
\put(835.0,849.0){\rule[-0.200pt]{0.400pt}{2.409pt}}
\put(900.0,163.0){\rule[-0.200pt]{0.400pt}{2.409pt}}
\put(900.0,849.0){\rule[-0.200pt]{0.400pt}{2.409pt}}
\put(955.0,163.0){\rule[-0.200pt]{0.400pt}{2.409pt}}
\put(955.0,849.0){\rule[-0.200pt]{0.400pt}{2.409pt}}
\put(1003.0,163.0){\rule[-0.200pt]{0.400pt}{2.409pt}}
\put(1003.0,849.0){\rule[-0.200pt]{0.400pt}{2.409pt}}
\put(1045.0,163.0){\rule[-0.200pt]{0.400pt}{2.409pt}}
\put(1045.0,849.0){\rule[-0.200pt]{0.400pt}{2.409pt}}
\put(1083.0,163.0){\rule[-0.200pt]{0.400pt}{4.818pt}}
\put(1083,122){\makebox(0,0){100}}
\put(1083.0,839.0){\rule[-0.200pt]{0.400pt}{4.818pt}}
\put(1330.0,163.0){\rule[-0.200pt]{0.400pt}{2.409pt}}
\put(1330.0,849.0){\rule[-0.200pt]{0.400pt}{2.409pt}}
\put(181.0,163.0){\rule[-0.200pt]{276.794pt}{0.400pt}}
\put(1330.0,163.0){\rule[-0.200pt]{0.400pt}{167.666pt}}
\put(181.0,859.0){\rule[-0.200pt]{276.794pt}{0.400pt}}
\put(41,511){\makebox(0,0){${{\langle E_{\rm k}\rangle}\over{L^2\mu^4}}$}}
\put(755,61){\makebox(0,0){$L$}}
\put(181.0,163.0){\rule[-0.200pt]{0.400pt}{167.666pt}}
\put(261,165){\raisebox{-.8pt}{\makebox(0,0){$\Diamond$}}}
\put(508,277){\raisebox{-.8pt}{\makebox(0,0){$\Diamond$}}}
\put(756,419){\raisebox{-.8pt}{\makebox(0,0){$\Diamond$}}}
\put(1003,592){\raisebox{-.8pt}{\makebox(0,0){$\Diamond$}}}
\put(1250,727){\raisebox{-.8pt}{\makebox(0,0){$\Diamond$}}}
\sbox{\plotpoint}{\rule[-0.400pt]{0.800pt}{0.800pt}}%
\put(261,165){\usebox{\plotpoint}}
\multiput(261.00,166.41)(1.106,0.501){217}{\rule{1.964pt}{0.121pt}}
\multiput(261.00,163.34)(242.923,112.000){2}{\rule{0.982pt}{0.800pt}}
\multiput(508.00,278.41)(0.811,0.501){299}{\rule{1.497pt}{0.121pt}}
\multiput(508.00,275.34)(244.893,153.000){2}{\rule{0.748pt}{0.800pt}}
\multiput(756.00,431.41)(0.637,0.500){381}{\rule{1.219pt}{0.121pt}}
\multiput(756.00,428.34)(244.471,194.000){2}{\rule{0.609pt}{0.800pt}}
\multiput(1003.00,625.41)(0.525,0.500){463}{\rule{1.041pt}{0.121pt}}
\multiput(1003.00,622.34)(244.840,235.000){2}{\rule{0.520pt}{0.800pt}}
\sbox{\plotpoint}{\rule[-0.500pt]{1.000pt}{1.000pt}}%
\sbox{\plotpoint}{\rule[-0.600pt]{1.200pt}{1.200pt}}%
\put(261,163){\makebox(0,0){$+$}}
\put(508,274){\makebox(0,0){$+$}}
\put(756,426){\makebox(0,0){$+$}}
\put(1003,612){\makebox(0,0){$+$}}
\put(1250,817){\makebox(0,0){$+$}}
\end{Large}
\end{picture}
\vskip-.6cm
\caption{The lattice size dependence of the scalar kinetic energy density,
expressed in units of $\mu^4$ for $\mu=0.025$ (pluses) and $\mu=0.05$
(diamonds). The solid line is the LPTh prediction. The error bars are smaller
than the plotting symbols.}
\label{ekvsl}
\end{figure}

The initial conditions for the fields $\alpha(\tau,r_t)$ and $\alpha_\perp^i$
at $\tau =0$ are obtained by matching the equations of motion
at the point $x^\pm =0$ and along the boundaries
$x^+=0,x^->0$ and $x^-=0,x^+>0$. Because the sources are highly singular
functions along their respective light cones, so too in general will be
 the equations of
motion. Remarkably, there exist a set of non--singular initial
conditions for the smooth evolution of the classical 
fields in the forward light
cone. In terms of the fields of each of the nuclei
before the collision ($t<0$), they are, 
$\alpha_\perp^i|_{\tau=0}= \alpha_1^i+\alpha_2^i$ and 
$\alpha|_{\tau=0}={ig\over 2} [\alpha_1^i,\alpha_2^i]$.
Gyulassy and McLerran have shown~\cite{gyulassy} that even when 
$\alpha_{1,2}^i$ are smeared out in rapidity to 
properly account for singular contact terms in the equations of motion, the
above boundary conditions remain unchanged.
Further, since the equations are very singular at $\tau=0$, the
only condition on the derivatives of the fields that leads to regular
solutions are $\partial_\tau \alpha|_{\tau=0},\partial_\tau \alpha_\perp^i
|_{\tau=0} =0$.

In Ref.~\cite{KLW}, perturbative solutions (for small $\rho$) were found 
to order $\rho^2$ by expanding 
the initial conditions and the fields in powers $\rho$ (or equivalently,
in powers of $\alpha_S\mu/k_t$) as  
$\alpha = \sum_{n=0}^{\infty} \alpha_{(n)}$\,\, ; \,\, 
$\alpha_{\perp}^i 
= \sum_{n=0}^{\infty} \alpha_{\perp (n)}^i$.
The subscript $n$ denotes the $n$th order in $\rho$. For
details of the perturbative solution, we refer the reader 
to their paper (see also the related paper of Gyulassy and 
McLerran~\cite{gyulassy}).

The perturbative Yang--Mills result for the number distribution of 
classical gluons is
\be
{dN\over {dyd^2 k_t}} = \pi R^2 {2g^6 \mu^4\over {(2\pi)^4}} {N_c (N_c^2-1)
\over k_t^4} L(k_t,\lambda) \, ,
\label{GunBer}
\ee
where $L(k_t,\lambda)$ is an infrared divergent function at the scale
$\lambda$. We first note that this result
agrees with the quantum bremsstrahlung formula of Gunion and
Bertsch~\cite{GunionBertsch} and with several later
works~\cite{gyulassy,DirkYuri,SerBerDir}.  It was also shown by Gyulassy and
McLerran that for sources smeared out in rapidity, the resulting 
expression is identical to the one above except $\mu^4\rightarrow
\chi^+(y)\chi^-(y)$. The $\pm$ superscripts refer to the nucleus on the
positive or negative light cone respectively.

The origin of the infrared divergent function $L(k_t,\lambda)$ above is from
long range color correlations that are cut-off either by a nuclear form
factor (as in Refs.~\cite{GunionBertsch,DirkYuri}), by dynamical screening
effects~\cite{GyuWang,EskMullWang} or in the classical Yang--Mills case of
Ref.~\cite{KLW}, non--linearities that become large at the scale $k_t\sim
\alpha_S\mu$. In the classical case then, $L(k_t,\lambda) =
\log(k_t^2/\lambda^2)$ where $\lambda = \alpha_S\mu$. 
At sufficiently high energies, the behaviour of
gluon radiation infrared is given by higher order (in
$\alpha_S\mu/k_t$) non--linear terms in the classical effective theory. 

It is unlikely that a simple analytical solution exists for
the Yang--Mills equations in general. The classical solutions then have to be
determined numerically for $t>0$. The straightforward procedure would be to
discretize the Yang--Mills equations
 but it will be more convenient for our purposes
to construct the lattice Hamiltonian and obtain the lattice equations of
motion from Hamilton's equations. This will be done in the next
section. Before we do that, we will discuss here the form of the continuum
Hamiltonian.

\begin{figure}[ht]
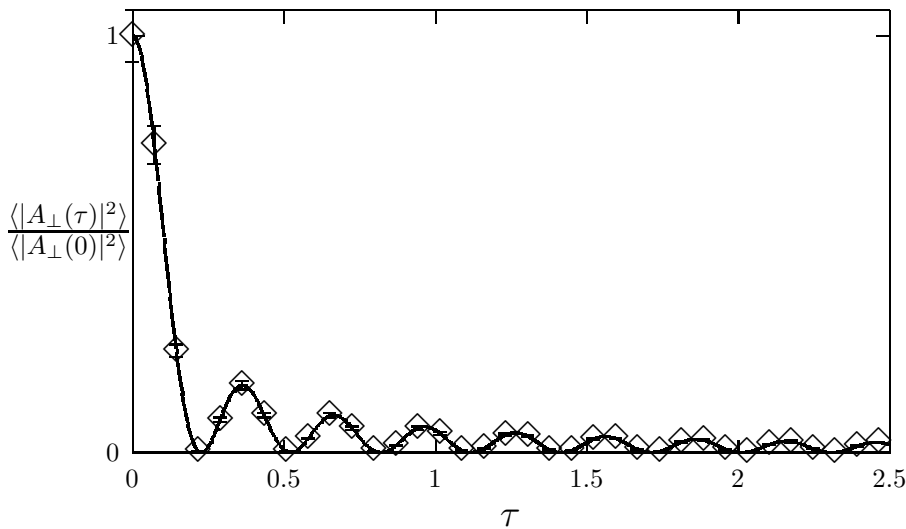

\setlength{\unitlength}{0.240900pt}
\ifx\plotpoint\undefined\newsavebox{\plotpoint}\fi
\sbox{\plotpoint}{\rule[-0.200pt]{0.400pt}{0.400pt}}%

\vskip-0.6cm
\caption{Normalized field intensity of a hard ($k_t=2.16{\rm GeV}$) mode 
vs proper time $\tau$ in units of fm (diamonds). Solid line is the LPTh 
prediction.}
\label{a2kvspht025pil160mu025}
\end{figure}

We start from the QCD action (without dynamical quarks) 
In the forward light cone ($t>0$) it is 
convenient to work with the $\tau,\eta,\vec{r_t}$ co--ordinates where 
$\tau=\sqrt{2 x^+ x^-}$ is the proper 
time, $\eta={1\over 2}\log(x^+/x^-)$ is the 
space--time rapidity and $\vec{r_t}=(x,y)$ are the two transverse Euclidean 
co--ordinates. In these co--ordinates, the metric is diagonal with 
$g^{\tau\tau}=-g^{xx}=-g^{yy}=1$ and $g^{\eta\eta}=-1/\tau^2$.

After a little algebra, the Hamiltonian can be written as~\cite{Sasha}
\be
H =\int d\eta d\vec{r_t} \tau \left\{{1\over 2} p^{\eta}p^{\eta}
+{1\over {2\tau^2}}p^r p^r + {1\over{2\tau^2}} F_{\eta r}F_{\eta r}
+{1\over 4}F_{xy}F_{xy} + j^\eta A_\eta + j^r A_r\right\} \, .
\label{hamilton}
\ee
Here we have adopted the Schwinger gauge condition which is 
equivalent to requiring $A^\tau =0$. 
Also, $p^\eta={1\over \tau}\partial_\tau A_\eta$ and $p^r=\tau \partial_\tau 
A_r$ are the conjugate momenta.

Consider the field strength $F_{\eta r}$ in the above Hamiltonian. If we 
assume approximate boost invariance, or
$A_r (\tau,\eta,\vec{r_t})\approx A_r(\tau,\vec{r_t})$ \, ; \,
$A_{\eta}(\tau,\eta,\vec{r_t})\approx \Phi(\tau,\vec{r_t})$,
we obtain  $F_{\eta r}^a = -D_r \Phi^a$.
Here $D_r =\partial_r -ig A_r$ is the covariant derivative. Further, if we 
express $j^{\eta,r}$ in terms of the $j^{\pm}$ defined in 
Eq.~\ref{sources} we obtain the result that $j^{\eta,r}=0$ for $\tau>0$.
Finally, since 
$\Phi = \tau^2 \alpha(\tau,\vec{r_t})$ \,; \, $A_r = \alpha_{\perp}^r 
(\tau, \vec{r_t})$,
we can perform the integration over the space--time rapidity to re--write
the Hamiltonian in Eq.~\ref{hamilton} as 
\be
H = \int d\vec{r_t} \tau \eta \left\{ {1\over 2}\left({\partial 
\alpha_{\perp}^r\over
{\partial\tau}}\right)^2 + {1\over 4}F_{xy}^a F_{xy}^a + {1\over {2\tau^2}}
\left(D_\beta \left[\tau^2 \alpha\right]\right)^2\right\} \, .
\label{twodh}
\ee
Here the index $\beta=(\tau,\vec{r_t})$. 
The discrete version of the
above Hamiltonian is the Kogut--Susskind
Hamiltonian~\cite{KS} in 2+1--dimensions coupled to an adjoint scalar field.
The lattice Hamiltonian will be discussed further in the next section.

We now briefly comment on a key assumption in the above derivation,
namely, the boost invariance of the fields.  This invariance results in
Eq.~\ref{twodh} thereby allowing us to restrict ourselves to a transverse
lattice alone. There has been some confusion about the importance of 
boost invariance as a consequence of the initial assumption that the
classical currents are delta functions on the light cone.  For remarks
clarifying this issue, see ref.~\cite{RajKrasnitz,RajKrasnitz2}.
In general, at the energies of interest, particle distributions are unlikely
to be boost invariant~\cite{KKV}.

\begin{figure}[ht]
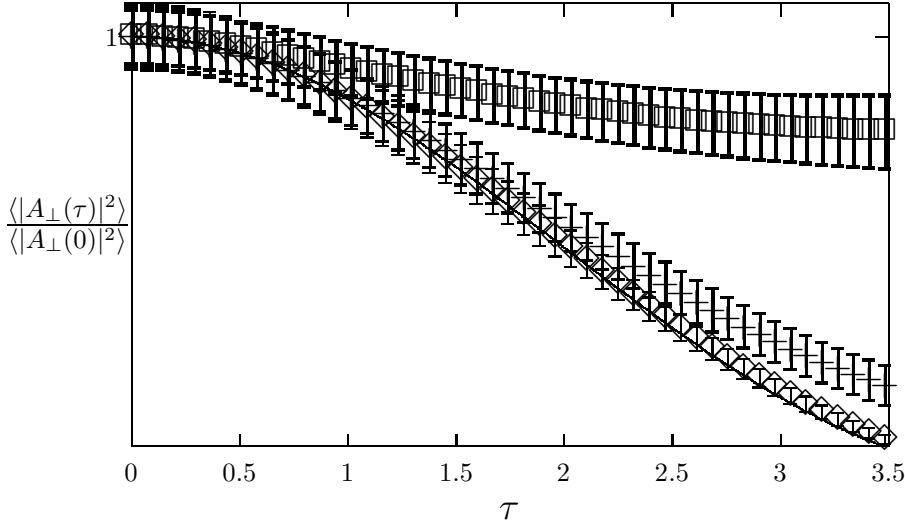

\setlength{\unitlength}{0.240900pt}
\ifx\plotpoint\undefined\newsavebox{\plotpoint}\fi
\sbox{\plotpoint}{\rule[-0.200pt]{0.400pt}{0.400pt}}%

\vskip-0.6cm
\caption{Normalized field intensity of a soft ($k_t=108{\rm MeV}$) mode 
vs proper time $\tau$ (in units of fm) for $\mu=200{\rm MeV}$ (squares),
$\mu=100{\rm MeV}$ (pluses), and $\mu=50{\rm MeV}$ (diamonds). Solid line, 
nearly coinciding with the $\mu=50{\rm MeV}$ curve, is the LPTh prediction.}
\label{a2kvsphtminkl160}
\end{figure}

\section{Real time lattice description of nuclear collisions}
\vspace*{0.3cm}

In this section, we discuss the problem on nuclear collisions on the lattice.
Taking the naive continuum limit in the longitudinal directions of the full 4
dimensional Minkowski Wilson action for the $SU(N_c)$ gauge group in
fundamental representation, one obtains an expression for the action $S$,
which is discrete in the transverse directions and continuous in the z--t
directions~\cite{RajKrasnitz2}.  The equation of motion for a field is
obtained by varying $S$ with respect to that field.  Initial conditions on
the lattice are derived from the lattice analogue of the continuum initial
conditions.  We start from the lattice action $S$, obtain the lattice
equations of motion in the four light cone regions and determine
non--singular initial conditions by matching at $\tau=0$ the coefficients of
the most singular terms in the equations of motion. The full derivation is
described in Ref.~\cite{RajKrasnitz2}.

The initial condition (at $\tau=0$) for the transverse link field for the
SU(2) case is
\be
U={{{\rm Tr}(U_1+U_2)}\over{U_1^\dagger+U_2^\dagger}}-I
=(U_1+U_2)(U_1^\dagger+U_2^\dagger)^{-1} \, .
\label{ucondp}
\ee
For $N_c>2$ the solution for $U$ is not as simple. This condition has the 
right formal continuum limit. This can be seen by writing $U_{1,2}$ 
as $\exp(ia_\perp \alpha_{1,2})$ and $U$ as $\exp(i a_\perp \alpha_\perp)$.
One obtains $\alpha_\perp = \alpha_1 + \alpha_2$ as required.

The result for the rapidity component of the gauge field (recall that
we had $A^\pm = \pm x^\pm \alpha (x_t,\tau)$) is
\bea
\alpha_\gamma&=&{i\over{4N_c}}\sum_n{\rm Tr}\sigma_\gamma
\Big([(U_1-U_2)(U^\dagger-I)-{\rm h.c.}]_{j,n}\nonumber \\
&-&[(U^\dagger-I)(U_1-U_2)-{\rm h.c.}]_{j-n,n}\Big)\, .
\label{agamma}
\eea
As above, in the limit of smooth fields, one obtains as required
$\alpha=i\sum_n[\alpha_1,\alpha_2]_n$.

The lattice Hamiltonian is obtained 
from the action following the standard Kogut-Susskind 
procedure~\cite{KS}, giving
\bea
H_L&=& {1\over{2\tau}}\sum_{l\equiv (j,n)} 
E_l^{a} E_l^{a} + \tau\sum_{\Box} \left(1-
{1\over 2}{\rm Tr} U_{\Box}\right)  \, ,\nonumber \\
&+& {1\over{4\tau}}\sum_{j,n}{\rm Tr}\,
\left(\Phi_j-U_{j,n}\Phi_{j+n}
U_{j,n}^\dagger\right)^2 +{\tau\over 4}\sum_j {\rm Tr}\,p_j^2,
\label{hl}
\eea
where $E_l$ are generators of right covariant derivatives on the group
and $U_{j,n}$ is a component of the usual SU(2) matrices corresponding
to a link from the site $j$ in the direction $n$. The first two terms
correspond to the contributions to the Hamiltonian from the chromoelectric and
chromomagnetic field strengths respectively. In the last equation
$\Phi\equiv \Phi^a\sigma^a$ is the adjoint scalar field with its conjugate
momentum $p\equiv p^a\sigma^a$. Matching the equations of motion at $\tau=0$
gives us the initial conditions $E_l|_{\tau=0} =0$, $\Phi_j|_{\tau=0}=0$ and 
$p_j|_{\tau=0} =2 \alpha$.

\begin{figure}[ht]
\setlength{\unitlength}{0.240900pt}
\ifx\plotpoint\undefined\newsavebox{\plotpoint}\fi
\sbox{\plotpoint}{\rule[-0.200pt]{0.400pt}{0.400pt}}%
\begin{picture}(1350,900)(0,0)
\begin{Large}
\font\gnuplot=cmr10 at 10pt
\gnuplot
\sbox{\plotpoint}{\rule[-0.200pt]{0.400pt}{0.400pt}}%
\put(241.0,466.0){\rule[-0.200pt]{4.818pt}{0.400pt}}
\put(221,466){\makebox(0,0)[r]{$10^7$}}
\put(1310.0,466.0){\rule[-0.200pt]{4.818pt}{0.400pt}}
\put(241.0,798.0){\rule[-0.200pt]{4.818pt}{0.400pt}}
\put(221,798){\makebox(0,0)[r]{$10^9$}}
\put(1310.0,798.0){\rule[-0.200pt]{4.818pt}{0.400pt}}
\put(241.0,163.0){\rule[-0.200pt]{0.400pt}{4.818pt}}
\put(241,122){\makebox(0,0){0}}
\put(241.0,839.0){\rule[-0.200pt]{0.400pt}{4.818pt}}
\put(489.0,163.0){\rule[-0.200pt]{0.400pt}{4.818pt}}
\put(489,122){\makebox(0,0){0.5}}
\put(489.0,839.0){\rule[-0.200pt]{0.400pt}{4.818pt}}
\put(736.0,163.0){\rule[-0.200pt]{0.400pt}{4.818pt}}
\put(736,122){\makebox(0,0){1}}
\put(736.0,839.0){\rule[-0.200pt]{0.400pt}{4.818pt}}
\put(983.0,163.0){\rule[-0.200pt]{0.400pt}{4.818pt}}
\put(983,122){\makebox(0,0){1.5}}
\put(983.0,839.0){\rule[-0.200pt]{0.400pt}{4.818pt}}
\put(1231.0,163.0){\rule[-0.200pt]{0.400pt}{4.818pt}}
\put(1231,122){\makebox(0,0){2}}
\put(1231.0,839.0){\rule[-0.200pt]{0.400pt}{4.818pt}}
\put(241.0,163.0){\rule[-0.200pt]{262.340pt}{0.400pt}}
\put(1330.0,163.0){\rule[-0.200pt]{0.400pt}{167.666pt}}
\put(241.0,859.0){\rule[-0.200pt]{262.340pt}{0.400pt}}
\put(80,511){\makebox(0,0){${{\langle |A_\perp|^2\rangle}\over{\mu^4}}$}}
\put(785,61){\makebox(0,0){$k_t$}}
\put(241.0,163.0){\rule[-0.200pt]{0.400pt}{167.666pt}}
\put(241,859){\raisebox{-.8pt}{\makebox(0,0){$\Diamond$}}}
\put(295,818){\raisebox{-.8pt}{\makebox(0,0){$\Diamond$}}}
\put(348,743){\raisebox{-.8pt}{\makebox(0,0){$\Diamond$}}}
\put(402,672){\raisebox{-.8pt}{\makebox(0,0){$\Diamond$}}}
\put(455,599){\raisebox{-.8pt}{\makebox(0,0){$\Diamond$}}}
\put(509,544){\raisebox{-.8pt}{\makebox(0,0){$\Diamond$}}}
\put(563,501){\raisebox{-.8pt}{\makebox(0,0){$\Diamond$}}}
\put(616,466){\raisebox{-.8pt}{\makebox(0,0){$\Diamond$}}}
\put(670,428){\raisebox{-.8pt}{\makebox(0,0){$\Diamond$}}}
\put(723,398){\raisebox{-.8pt}{\makebox(0,0){$\Diamond$}}}
\put(777,373){\raisebox{-.8pt}{\makebox(0,0){$\Diamond$}}}
\put(831,351){\raisebox{-.8pt}{\makebox(0,0){$\Diamond$}}}
\put(884,323){\raisebox{-.8pt}{\makebox(0,0){$\Diamond$}}}
\put(938,306){\raisebox{-.8pt}{\makebox(0,0){$\Diamond$}}}
\put(991,282){\raisebox{-.8pt}{\makebox(0,0){$\Diamond$}}}
\put(1045,264){\raisebox{-.8pt}{\makebox(0,0){$\Diamond$}}}
\put(1098,243){\raisebox{-.8pt}{\makebox(0,0){$\Diamond$}}}
\put(1152,231){\raisebox{-.8pt}{\makebox(0,0){$\Diamond$}}}
\put(1206,220){\raisebox{-.8pt}{\makebox(0,0){$\Diamond$}}}
\put(1259,204){\raisebox{-.8pt}{\makebox(0,0){$\Diamond$}}}
\put(1313,188){\raisebox{-.8pt}{\makebox(0,0){$\Diamond$}}}
\sbox{\plotpoint}{\rule[-0.400pt]{0.800pt}{0.800pt}}%
\put(241,822){\makebox(0,0){$+$}}
\put(295,790){\makebox(0,0){$+$}}
\put(348,731){\makebox(0,0){$+$}}
\put(402,670){\makebox(0,0){$+$}}
\put(455,604){\makebox(0,0){$+$}}
\put(509,553){\makebox(0,0){$+$}}
\put(563,494){\makebox(0,0){$+$}}
\put(616,459){\makebox(0,0){$+$}}
\put(670,428){\makebox(0,0){$+$}}
\put(723,396){\makebox(0,0){$+$}}
\put(777,378){\makebox(0,0){$+$}}
\put(831,335){\makebox(0,0){$+$}}
\put(884,317){\makebox(0,0){$+$}}
\put(938,303){\makebox(0,0){$+$}}
\put(991,283){\makebox(0,0){$+$}}
\put(1045,264){\makebox(0,0){$+$}}
\put(1098,244){\makebox(0,0){$+$}}
\put(1152,227){\makebox(0,0){$+$}}
\put(1206,205){\makebox(0,0){$+$}}
\put(1259,200){\makebox(0,0){$+$}}
\put(1313,181){\makebox(0,0){$+$}}
\sbox{\plotpoint}{\rule[-0.500pt]{1.000pt}{1.000pt}}%
\sbox{\plotpoint}{\rule[-0.600pt]{1.200pt}{1.200pt}}%
\put(241,684){\raisebox{-.8pt}{\makebox(0,0){$\Box$}}}
\put(295,674){\raisebox{-.8pt}{\makebox(0,0){$\Box$}}}
\put(348,645){\raisebox{-.8pt}{\makebox(0,0){$\Box$}}}
\put(402,607){\raisebox{-.8pt}{\makebox(0,0){$\Box$}}}
\put(455,560){\raisebox{-.8pt}{\makebox(0,0){$\Box$}}}
\put(509,524){\raisebox{-.8pt}{\makebox(0,0){$\Box$}}}
\put(563,474){\raisebox{-.8pt}{\makebox(0,0){$\Box$}}}
\put(616,443){\raisebox{-.8pt}{\makebox(0,0){$\Box$}}}
\put(670,404){\raisebox{-.8pt}{\makebox(0,0){$\Box$}}}
\put(723,374){\raisebox{-.8pt}{\makebox(0,0){$\Box$}}}
\put(777,347){\raisebox{-.8pt}{\makebox(0,0){$\Box$}}}
\put(831,320){\raisebox{-.8pt}{\makebox(0,0){$\Box$}}}
\put(884,289){\raisebox{-.8pt}{\makebox(0,0){$\Box$}}}
\put(938,268){\raisebox{-.8pt}{\makebox(0,0){$\Box$}}}
\put(991,265){\raisebox{-.8pt}{\makebox(0,0){$\Box$}}}
\put(1045,243){\raisebox{-.8pt}{\makebox(0,0){$\Box$}}}
\put(1098,228){\raisebox{-.8pt}{\makebox(0,0){$\Box$}}}
\put(1152,205){\raisebox{-.8pt}{\makebox(0,0){$\Box$}}}
\put(1206,188){\raisebox{-.8pt}{\makebox(0,0){$\Box$}}}
\put(1259,179){\raisebox{-.8pt}{\makebox(0,0){$\Box$}}}
\put(1313,163){\raisebox{-.8pt}{\makebox(0,0){$\Box$}}}
\sbox{\plotpoint}{\rule[-0.500pt]{1.000pt}{1.000pt}}%
\sbox{\plotpoint}{\rule[-0.200pt]{0.400pt}{0.400pt}}%
\put(295,840){\usebox{\plotpoint}}
\multiput(295.58,836.92)(0.498,-0.803){103}{\rule{0.120pt}{0.742pt}}
\multiput(294.17,838.46)(53.000,-83.461){2}{\rule{0.400pt}{0.371pt}}
\multiput(348.58,752.06)(0.498,-0.760){105}{\rule{0.120pt}{0.707pt}}
\multiput(347.17,753.53)(54.000,-80.532){2}{\rule{0.400pt}{0.354pt}}
\multiput(402.58,670.49)(0.498,-0.632){103}{\rule{0.120pt}{0.606pt}}
\multiput(401.17,671.74)(53.000,-65.743){2}{\rule{0.400pt}{0.303pt}}
\multiput(455.58,603.89)(0.498,-0.509){105}{\rule{0.120pt}{0.507pt}}
\multiput(454.17,604.95)(54.000,-53.947){2}{\rule{0.400pt}{0.254pt}}
\multiput(509.00,549.92)(0.600,-0.498){87}{\rule{0.580pt}{0.120pt}}
\multiput(509.00,550.17)(52.796,-45.000){2}{\rule{0.290pt}{0.400pt}}
\multiput(563.00,504.92)(0.680,-0.498){75}{\rule{0.644pt}{0.120pt}}
\multiput(563.00,505.17)(51.664,-39.000){2}{\rule{0.322pt}{0.400pt}}
\multiput(616.00,465.92)(0.773,-0.498){67}{\rule{0.717pt}{0.120pt}}
\multiput(616.00,466.17)(52.512,-35.000){2}{\rule{0.359pt}{0.400pt}}
\multiput(670.00,430.92)(0.887,-0.497){57}{\rule{0.807pt}{0.120pt}}
\multiput(670.00,431.17)(51.326,-30.000){2}{\rule{0.403pt}{0.400pt}}
\multiput(723.00,400.92)(1.005,-0.497){51}{\rule{0.900pt}{0.120pt}}
\multiput(723.00,401.17)(52.132,-27.000){2}{\rule{0.450pt}{0.400pt}}
\multiput(777.00,373.92)(1.087,-0.497){47}{\rule{0.964pt}{0.120pt}}
\multiput(777.00,374.17)(51.999,-25.000){2}{\rule{0.482pt}{0.400pt}}
\multiput(831.00,348.92)(1.215,-0.496){41}{\rule{1.064pt}{0.120pt}}
\multiput(831.00,349.17)(50.792,-22.000){2}{\rule{0.532pt}{0.400pt}}
\multiput(884.00,326.92)(1.298,-0.496){39}{\rule{1.129pt}{0.119pt}}
\multiput(884.00,327.17)(51.658,-21.000){2}{\rule{0.564pt}{0.400pt}}
\multiput(938.00,305.92)(1.410,-0.495){35}{\rule{1.216pt}{0.119pt}}
\multiput(938.00,306.17)(50.477,-19.000){2}{\rule{0.608pt}{0.400pt}}
\multiput(991.00,286.92)(1.519,-0.495){33}{\rule{1.300pt}{0.119pt}}
\multiput(991.00,287.17)(51.302,-18.000){2}{\rule{0.650pt}{0.400pt}}
\multiput(1045.00,268.92)(1.682,-0.494){29}{\rule{1.425pt}{0.119pt}}
\multiput(1045.00,269.17)(50.042,-16.000){2}{\rule{0.712pt}{0.400pt}}
\multiput(1098.00,252.92)(1.714,-0.494){29}{\rule{1.450pt}{0.119pt}}
\multiput(1098.00,253.17)(50.990,-16.000){2}{\rule{0.725pt}{0.400pt}}
\multiput(1152.00,236.92)(1.831,-0.494){27}{\rule{1.540pt}{0.119pt}}
\multiput(1152.00,237.17)(50.804,-15.000){2}{\rule{0.770pt}{0.400pt}}
\multiput(1206.00,221.92)(1.929,-0.494){25}{\rule{1.614pt}{0.119pt}}
\multiput(1206.00,222.17)(49.649,-14.000){2}{\rule{0.807pt}{0.400pt}}
\multiput(1259.00,207.92)(2.122,-0.493){23}{\rule{1.762pt}{0.119pt}}
\multiput(1259.00,208.17)(50.344,-13.000){2}{\rule{0.881pt}{0.400pt}}
\multiput(1313.00,194.94)(2.382,-0.468){5}{\rule{1.800pt}{0.113pt}}
\multiput(1313.00,195.17)(13.264,-4.000){2}{\rule{0.900pt}{0.400pt}}
\end{Large}
\end{picture}
\vskip-0.6cm
\caption{Field intensity over $\mu^4$ as a function of $k_t$ 
for $\mu=200{\rm MeV}$ (squares),
$\mu=100{\rm MeV}$ (pluses), and $\mu=50{\rm MeV}$ (diamonds). Solid line 
is the LPTh prediction. The field intensity is in arbitrary units and
$k_t$ is in GeV.}
\label{na2vskl160}
\end{figure}
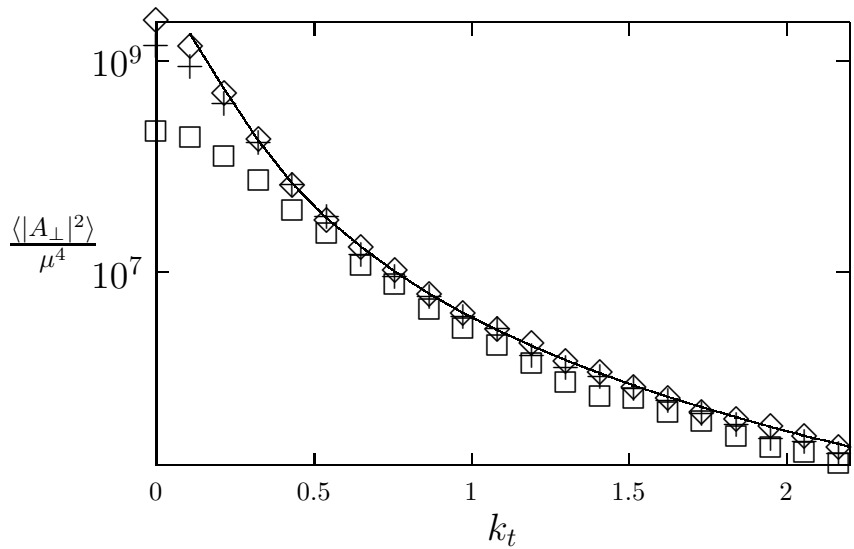

Lattice equations of motion follow directly from $H_L$ of Eq.~\ref{hl}.  For
any dynamical variable $v$ with no explicit time dependence, ${\dot
v}=\{H_L,v\}$, where ${\dot v}$ is the derivative with respect to $\tau$, and
$\{\}$ denote Poisson brackets. We take $E_l$, $U_l$, $p_j$, and $\Phi_j$ as
independent dynamical variables, whose only nonvanishing Poisson brackets are
$$\{p_i^a,\Phi_j^b\}=\delta_{ij}\delta_{ab}; \ \ 
\{E_l^a,U_m\}=-i\delta_{lm}U_l\sigma^a; \ \
\{E_l^a,E_m^b\}=2\delta_{lm}\epsilon_{abc}E_l^c$$
(no summing of repeated indices). The equations of motion are consistent with
a set of local constraints (Gauss' laws). 

\section{Mini--jets on the lattice: numerical results and comparison to 
lattice perturbation theory}
\vspace*{0.3cm} 

In this section, we will compare results from our numerical simulations to
analytic results from lattice perturbation theory. Before we
do that we should first understand the ramifications of our results 
for nuclear collisions.

In section 2, we introduced the scale $\mu^2$, which is the color charge 
squared per unit area.  Its magnitude can be determined from the
relation~\cite{gyulassy}
\be
\mu^2 = {A^{1/3}\over {\pi r_0^2}} \int_{x_0}^1 dx \left({1\over 2 N_c} 
q(x,Q^2) + {N_c\over {N_c^2-1}} g(x,Q^2)\right) \, ,
\ee
where $q,g$ stand for the {\it nucleon} 
quark and gluon structure functions at the resolution 
scale $Q$ of the physical process of interest. Also, above $x_0 =
Q/\sqrt{s}$. Using the HERA structure function data, Gyulassy and McLerran
estimated that $\mu\leq 1$ GeV for LHC energies and $\mu \leq 0..5$ GeV 
at RHIC. Thus the regime where the classical Yang--Mills picture can be
applied is rather limited. However, 
a window of applicability 
does exist and depending on what higher order calculations will tell
us, this window may be larger or smaller than the naive classical extimates.

Gyulassy and McLerran~\cite{gyulassy} have shown that the classical 
Yang--Mills formula in Eq.~\ref{GunBer} is at small x (approximately) 
the same as the perturbative QCD prediction~\cite{GLR} for the process 
$AA\rightarrow g$ 
\be
{d\sigma\over {dyd^2 k_t}} = K_N {\alpha_S N_c\over {\pi^2 k_t^2}}
\int d^2 q_t {f(x_1,q_t^2) f(x_2,(\vec{k}_t-\vec{q}_t)^2)\over {q_t^2 
(\vec{k}_t-\vec{q}_t)^2}} \, ,
\ee
where 
$f(x,Q^2) = {d xG(x,Q^2) \over {d\log{Q^2}}}$,
and $x_1\approx x_2 = k_t/\sqrt{s}$. The two formulae are equivalent if we
divide the above formula by $\pi R^2$, approximate the integral above by
factoring out $f$ above at the scale $k_t^2$ and taking the normalization
factor $K_N\approx 5$.

At large transverse momenta, 
the field intensity measured on the lattice, $|A (k_t,\tau)|^2$, is
simply related to the Yang--Mills distribution function. Then, from the
above discussion, the field intensity of the hard modes 
on the lattice can also be simply related to the 
mini--jet cross section for the process $AA\rightarrow g$.
However, at smaller transverse momenta, there is no simple relation between
the field intensity and the classical gluon distribution function (and
thereby the cross section by the above arguments). We have to look for more
general quantities which, conversely, in the limit of large $k_t$, will give
us the $AA\rightarrow g$ cross section. This point will be discussed further
in Ref.~\cite{RajKrasnitz2}.

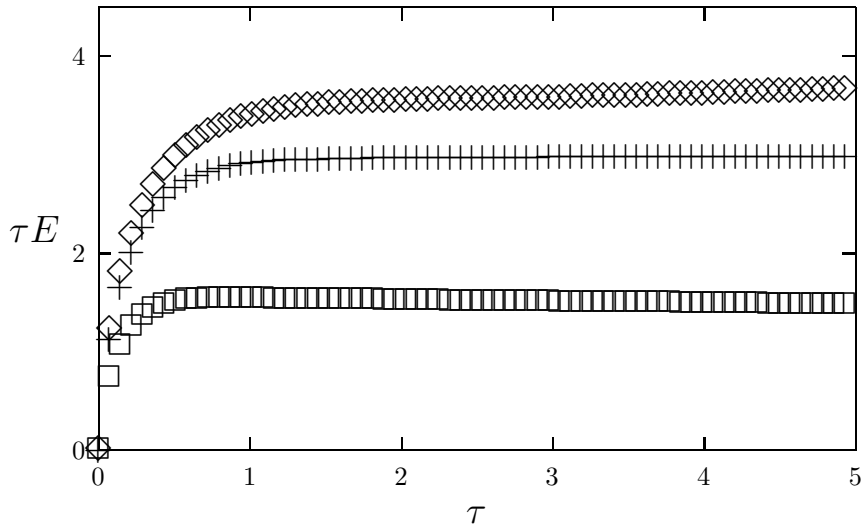
\begin{figure}[ht]
\setlength{\unitlength}{0.240900pt}
\ifx\plotpoint\undefined\newsavebox{\plotpoint}\fi
\sbox{\plotpoint}{\rule[-0.200pt]{0.400pt}{0.400pt}}%
\begin{picture}(1350,900)(0,0)
\begin{Large}
\font\gnuplot=cmr10 at 10pt
\gnuplot
\sbox{\plotpoint}{\rule[-0.200pt]{0.400pt}{0.400pt}}%
\put(141.0,163.0){\rule[-0.200pt]{4.818pt}{0.400pt}}
\put(121,163){\makebox(0,0)[r]{0}}
\put(1310.0,163.0){\rule[-0.200pt]{4.818pt}{0.400pt}}
\put(141.0,472.0){\rule[-0.200pt]{4.818pt}{0.400pt}}
\put(121,472){\makebox(0,0)[r]{2}}
\put(1310.0,472.0){\rule[-0.200pt]{4.818pt}{0.400pt}}
\put(141.0,782.0){\rule[-0.200pt]{4.818pt}{0.400pt}}
\put(121,782){\makebox(0,0)[r]{4}}
\put(1310.0,782.0){\rule[-0.200pt]{4.818pt}{0.400pt}}
\put(141.0,163.0){\rule[-0.200pt]{0.400pt}{4.818pt}}
\put(141,122){\makebox(0,0){0}}
\put(141.0,839.0){\rule[-0.200pt]{0.400pt}{4.818pt}}
\put(379.0,163.0){\rule[-0.200pt]{0.400pt}{4.818pt}}
\put(379,122){\makebox(0,0){1}}
\put(379.0,839.0){\rule[-0.200pt]{0.400pt}{4.818pt}}
\put(617.0,163.0){\rule[-0.200pt]{0.400pt}{4.818pt}}
\put(617,122){\makebox(0,0){2}}
\put(617.0,839.0){\rule[-0.200pt]{0.400pt}{4.818pt}}
\put(854.0,163.0){\rule[-0.200pt]{0.400pt}{4.818pt}}
\put(854,122){\makebox(0,0){3}}
\put(854.0,839.0){\rule[-0.200pt]{0.400pt}{4.818pt}}
\put(1092.0,163.0){\rule[-0.200pt]{0.400pt}{4.818pt}}
\put(1092,122){\makebox(0,0){4}}
\put(1092.0,839.0){\rule[-0.200pt]{0.400pt}{4.818pt}}
\put(1330.0,163.0){\rule[-0.200pt]{0.400pt}{4.818pt}}
\put(1330,122){\makebox(0,0){5}}
\put(1330.0,839.0){\rule[-0.200pt]{0.400pt}{4.818pt}}
\put(141.0,163.0){\rule[-0.200pt]{286.430pt}{0.400pt}}
\put(1330.0,163.0){\rule[-0.200pt]{0.400pt}{167.666pt}}
\put(141.0,859.0){\rule[-0.200pt]{286.430pt}{0.400pt}}
\put(41,511){\makebox(0,0){$\tau E$}}
\put(735,61){\makebox(0,0){$\tau$}}
\put(141.0,163.0){\rule[-0.200pt]{0.400pt}{167.666pt}}
\put(141,163){\raisebox{-.8pt}{\makebox(0,0){$\Diamond$}}}
\put(158,351){\raisebox{-.8pt}{\makebox(0,0){$\Diamond$}}}
\put(175,441){\raisebox{-.8pt}{\makebox(0,0){$\Diamond$}}}
\put(193,501){\raisebox{-.8pt}{\makebox(0,0){$\Diamond$}}}
\put(210,545){\raisebox{-.8pt}{\makebox(0,0){$\Diamond$}}}
\put(227,578){\raisebox{-.8pt}{\makebox(0,0){$\Diamond$}}}
\put(244,603){\raisebox{-.8pt}{\makebox(0,0){$\Diamond$}}}
\put(262,623){\raisebox{-.8pt}{\makebox(0,0){$\Diamond$}}}
\put(279,639){\raisebox{-.8pt}{\makebox(0,0){$\Diamond$}}}
\put(296,652){\raisebox{-.8pt}{\makebox(0,0){$\Diamond$}}}
\put(313,662){\raisebox{-.8pt}{\makebox(0,0){$\Diamond$}}}
\put(331,671){\raisebox{-.8pt}{\makebox(0,0){$\Diamond$}}}
\put(348,678){\raisebox{-.8pt}{\makebox(0,0){$\Diamond$}}}
\put(365,684){\raisebox{-.8pt}{\makebox(0,0){$\Diamond$}}}
\put(382,688){\raisebox{-.8pt}{\makebox(0,0){$\Diamond$}}}
\put(400,692){\raisebox{-.8pt}{\makebox(0,0){$\Diamond$}}}
\put(417,696){\raisebox{-.8pt}{\makebox(0,0){$\Diamond$}}}
\put(434,698){\raisebox{-.8pt}{\makebox(0,0){$\Diamond$}}}
\put(451,701){\raisebox{-.8pt}{\makebox(0,0){$\Diamond$}}}
\put(469,703){\raisebox{-.8pt}{\makebox(0,0){$\Diamond$}}}
\put(486,704){\raisebox{-.8pt}{\makebox(0,0){$\Diamond$}}}
\put(503,706){\raisebox{-.8pt}{\makebox(0,0){$\Diamond$}}}
\put(520,707){\raisebox{-.8pt}{\makebox(0,0){$\Diamond$}}}
\put(538,708){\raisebox{-.8pt}{\makebox(0,0){$\Diamond$}}}
\put(555,708){\raisebox{-.8pt}{\makebox(0,0){$\Diamond$}}}
\put(572,709){\raisebox{-.8pt}{\makebox(0,0){$\Diamond$}}}
\put(589,710){\raisebox{-.8pt}{\makebox(0,0){$\Diamond$}}}
\put(606,710){\raisebox{-.8pt}{\makebox(0,0){$\Diamond$}}}
\put(624,711){\raisebox{-.8pt}{\makebox(0,0){$\Diamond$}}}
\put(641,711){\raisebox{-.8pt}{\makebox(0,0){$\Diamond$}}}
\put(658,711){\raisebox{-.8pt}{\makebox(0,0){$\Diamond$}}}
\put(675,712){\raisebox{-.8pt}{\makebox(0,0){$\Diamond$}}}
\put(693,712){\raisebox{-.8pt}{\makebox(0,0){$\Diamond$}}}
\put(710,712){\raisebox{-.8pt}{\makebox(0,0){$\Diamond$}}}
\put(727,713){\raisebox{-.8pt}{\makebox(0,0){$\Diamond$}}}
\put(744,713){\raisebox{-.8pt}{\makebox(0,0){$\Diamond$}}}
\put(762,713){\raisebox{-.8pt}{\makebox(0,0){$\Diamond$}}}
\put(779,714){\raisebox{-.8pt}{\makebox(0,0){$\Diamond$}}}
\put(796,714){\raisebox{-.8pt}{\makebox(0,0){$\Diamond$}}}
\put(813,714){\raisebox{-.8pt}{\makebox(0,0){$\Diamond$}}}
\put(831,715){\raisebox{-.8pt}{\makebox(0,0){$\Diamond$}}}
\put(848,715){\raisebox{-.8pt}{\makebox(0,0){$\Diamond$}}}
\put(865,715){\raisebox{-.8pt}{\makebox(0,0){$\Diamond$}}}
\put(882,716){\raisebox{-.8pt}{\makebox(0,0){$\Diamond$}}}
\put(900,716){\raisebox{-.8pt}{\makebox(0,0){$\Diamond$}}}
\put(917,717){\raisebox{-.8pt}{\makebox(0,0){$\Diamond$}}}
\put(934,717){\raisebox{-.8pt}{\makebox(0,0){$\Diamond$}}}
\put(951,717){\raisebox{-.8pt}{\makebox(0,0){$\Diamond$}}}
\put(969,718){\raisebox{-.8pt}{\makebox(0,0){$\Diamond$}}}
\put(986,718){\raisebox{-.8pt}{\makebox(0,0){$\Diamond$}}}
\put(1003,719){\raisebox{-.8pt}{\makebox(0,0){$\Diamond$}}}
\put(1020,719){\raisebox{-.8pt}{\makebox(0,0){$\Diamond$}}}
\put(1038,720){\raisebox{-.8pt}{\makebox(0,0){$\Diamond$}}}
\put(1055,720){\raisebox{-.8pt}{\makebox(0,0){$\Diamond$}}}
\put(1072,721){\raisebox{-.8pt}{\makebox(0,0){$\Diamond$}}}
\put(1089,721){\raisebox{-.8pt}{\makebox(0,0){$\Diamond$}}}
\put(1106,722){\raisebox{-.8pt}{\makebox(0,0){$\Diamond$}}}
\put(1124,722){\raisebox{-.8pt}{\makebox(0,0){$\Diamond$}}}
\put(1141,723){\raisebox{-.8pt}{\makebox(0,0){$\Diamond$}}}
\put(1158,723){\raisebox{-.8pt}{\makebox(0,0){$\Diamond$}}}
\put(1175,724){\raisebox{-.8pt}{\makebox(0,0){$\Diamond$}}}
\put(1193,725){\raisebox{-.8pt}{\makebox(0,0){$\Diamond$}}}
\put(1210,725){\raisebox{-.8pt}{\makebox(0,0){$\Diamond$}}}
\put(1227,726){\raisebox{-.8pt}{\makebox(0,0){$\Diamond$}}}
\put(1244,726){\raisebox{-.8pt}{\makebox(0,0){$\Diamond$}}}
\put(1262,727){\raisebox{-.8pt}{\makebox(0,0){$\Diamond$}}}
\put(1279,727){\raisebox{-.8pt}{\makebox(0,0){$\Diamond$}}}
\put(1296,728){\raisebox{-.8pt}{\makebox(0,0){$\Diamond$}}}
\put(1313,729){\raisebox{-.8pt}{\makebox(0,0){$\Diamond$}}}
\sbox{\plotpoint}{\rule[-0.400pt]{0.800pt}{0.800pt}}%
\put(141,163){\makebox(0,0){$+$}}
\put(158,336){\makebox(0,0){$+$}}
\put(175,419){\makebox(0,0){$+$}}
\put(193,473){\makebox(0,0){$+$}}
\put(210,512){\makebox(0,0){$+$}}
\put(227,539){\makebox(0,0){$+$}}
\put(244,560){\makebox(0,0){$+$}}
\put(262,575){\makebox(0,0){$+$}}
\put(279,586){\makebox(0,0){$+$}}
\put(296,595){\makebox(0,0){$+$}}
\put(313,601){\makebox(0,0){$+$}}
\put(331,606){\makebox(0,0){$+$}}
\put(348,610){\makebox(0,0){$+$}}
\put(365,613){\makebox(0,0){$+$}}
\put(382,615){\makebox(0,0){$+$}}
\put(400,616){\makebox(0,0){$+$}}
\put(417,618){\makebox(0,0){$+$}}
\put(434,619){\makebox(0,0){$+$}}
\put(451,619){\makebox(0,0){$+$}}
\put(469,620){\makebox(0,0){$+$}}
\put(486,620){\makebox(0,0){$+$}}
\put(503,621){\makebox(0,0){$+$}}
\put(520,621){\makebox(0,0){$+$}}
\put(538,621){\makebox(0,0){$+$}}
\put(555,621){\makebox(0,0){$+$}}
\put(572,622){\makebox(0,0){$+$}}
\put(589,622){\makebox(0,0){$+$}}
\put(606,622){\makebox(0,0){$+$}}
\put(624,622){\makebox(0,0){$+$}}
\put(641,622){\makebox(0,0){$+$}}
\put(658,622){\makebox(0,0){$+$}}
\put(675,622){\makebox(0,0){$+$}}
\put(693,622){\makebox(0,0){$+$}}
\put(710,622){\makebox(0,0){$+$}}
\put(727,622){\makebox(0,0){$+$}}
\put(744,623){\makebox(0,0){$+$}}
\put(762,623){\makebox(0,0){$+$}}
\put(779,623){\makebox(0,0){$+$}}
\put(796,623){\makebox(0,0){$+$}}
\put(813,623){\makebox(0,0){$+$}}
\put(831,623){\makebox(0,0){$+$}}
\put(848,624){\makebox(0,0){$+$}}
\put(865,624){\makebox(0,0){$+$}}
\put(882,624){\makebox(0,0){$+$}}
\put(900,624){\makebox(0,0){$+$}}
\put(917,624){\makebox(0,0){$+$}}
\put(934,624){\makebox(0,0){$+$}}
\put(951,624){\makebox(0,0){$+$}}
\put(969,625){\makebox(0,0){$+$}}
\put(986,625){\makebox(0,0){$+$}}
\put(1003,625){\makebox(0,0){$+$}}
\put(1020,625){\makebox(0,0){$+$}}
\put(1038,625){\makebox(0,0){$+$}}
\put(1055,625){\makebox(0,0){$+$}}
\put(1072,625){\makebox(0,0){$+$}}
\put(1089,625){\makebox(0,0){$+$}}
\put(1106,625){\makebox(0,0){$+$}}
\put(1124,625){\makebox(0,0){$+$}}
\put(1141,625){\makebox(0,0){$+$}}
\put(1158,625){\makebox(0,0){$+$}}
\put(1175,625){\makebox(0,0){$+$}}
\put(1193,625){\makebox(0,0){$+$}}
\put(1210,625){\makebox(0,0){$+$}}
\put(1227,625){\makebox(0,0){$+$}}
\put(1244,625){\makebox(0,0){$+$}}
\put(1262,625){\makebox(0,0){$+$}}
\put(1279,625){\makebox(0,0){$+$}}
\put(1296,625){\makebox(0,0){$+$}}
\put(1313,625){\makebox(0,0){$+$}}
\sbox{\plotpoint}{\rule[-0.500pt]{1.000pt}{1.000pt}}%
\sbox{\plotpoint}{\rule[-0.600pt]{1.200pt}{1.200pt}}%
\put(141,163){\raisebox{-.8pt}{\makebox(0,0){$\Box$}}}
\put(158,276){\raisebox{-.8pt}{\makebox(0,0){$\Box$}}}
\put(175,326){\raisebox{-.8pt}{\makebox(0,0){$\Box$}}}
\put(193,356){\raisebox{-.8pt}{\makebox(0,0){$\Box$}}}
\put(210,374){\raisebox{-.8pt}{\makebox(0,0){$\Box$}}}
\put(227,385){\raisebox{-.8pt}{\makebox(0,0){$\Box$}}}
\put(244,391){\raisebox{-.8pt}{\makebox(0,0){$\Box$}}}
\put(262,395){\raisebox{-.8pt}{\makebox(0,0){$\Box$}}}
\put(279,398){\raisebox{-.8pt}{\makebox(0,0){$\Box$}}}
\put(296,399){\raisebox{-.8pt}{\makebox(0,0){$\Box$}}}
\put(313,400){\raisebox{-.8pt}{\makebox(0,0){$\Box$}}}
\put(331,400){\raisebox{-.8pt}{\makebox(0,0){$\Box$}}}
\put(348,400){\raisebox{-.8pt}{\makebox(0,0){$\Box$}}}
\put(365,400){\raisebox{-.8pt}{\makebox(0,0){$\Box$}}}
\put(382,400){\raisebox{-.8pt}{\makebox(0,0){$\Box$}}}
\put(400,400){\raisebox{-.8pt}{\makebox(0,0){$\Box$}}}
\put(417,399){\raisebox{-.8pt}{\makebox(0,0){$\Box$}}}
\put(434,399){\raisebox{-.8pt}{\makebox(0,0){$\Box$}}}
\put(451,399){\raisebox{-.8pt}{\makebox(0,0){$\Box$}}}
\put(469,399){\raisebox{-.8pt}{\makebox(0,0){$\Box$}}}
\put(486,399){\raisebox{-.8pt}{\makebox(0,0){$\Box$}}}
\put(503,398){\raisebox{-.8pt}{\makebox(0,0){$\Box$}}}
\put(520,398){\raisebox{-.8pt}{\makebox(0,0){$\Box$}}}
\put(538,398){\raisebox{-.8pt}{\makebox(0,0){$\Box$}}}
\put(555,398){\raisebox{-.8pt}{\makebox(0,0){$\Box$}}}
\put(572,398){\raisebox{-.8pt}{\makebox(0,0){$\Box$}}}
\put(589,397){\raisebox{-.8pt}{\makebox(0,0){$\Box$}}}
\put(606,397){\raisebox{-.8pt}{\makebox(0,0){$\Box$}}}
\put(624,397){\raisebox{-.8pt}{\makebox(0,0){$\Box$}}}
\put(641,397){\raisebox{-.8pt}{\makebox(0,0){$\Box$}}}
\put(658,397){\raisebox{-.8pt}{\makebox(0,0){$\Box$}}}
\put(675,397){\raisebox{-.8pt}{\makebox(0,0){$\Box$}}}
\put(693,396){\raisebox{-.8pt}{\makebox(0,0){$\Box$}}}
\put(710,396){\raisebox{-.8pt}{\makebox(0,0){$\Box$}}}
\put(727,396){\raisebox{-.8pt}{\makebox(0,0){$\Box$}}}
\put(744,396){\raisebox{-.8pt}{\makebox(0,0){$\Box$}}}
\put(762,396){\raisebox{-.8pt}{\makebox(0,0){$\Box$}}}
\put(779,395){\raisebox{-.8pt}{\makebox(0,0){$\Box$}}}
\put(796,395){\raisebox{-.8pt}{\makebox(0,0){$\Box$}}}
\put(813,395){\raisebox{-.8pt}{\makebox(0,0){$\Box$}}}
\put(831,395){\raisebox{-.8pt}{\makebox(0,0){$\Box$}}}
\put(848,395){\raisebox{-.8pt}{\makebox(0,0){$\Box$}}}
\put(865,394){\raisebox{-.8pt}{\makebox(0,0){$\Box$}}}
\put(882,394){\raisebox{-.8pt}{\makebox(0,0){$\Box$}}}
\put(900,394){\raisebox{-.8pt}{\makebox(0,0){$\Box$}}}
\put(917,394){\raisebox{-.8pt}{\makebox(0,0){$\Box$}}}
\put(934,394){\raisebox{-.8pt}{\makebox(0,0){$\Box$}}}
\put(951,393){\raisebox{-.8pt}{\makebox(0,0){$\Box$}}}
\put(969,393){\raisebox{-.8pt}{\makebox(0,0){$\Box$}}}
\put(986,393){\raisebox{-.8pt}{\makebox(0,0){$\Box$}}}
\put(1003,393){\raisebox{-.8pt}{\makebox(0,0){$\Box$}}}
\put(1020,393){\raisebox{-.8pt}{\makebox(0,0){$\Box$}}}
\put(1038,393){\raisebox{-.8pt}{\makebox(0,0){$\Box$}}}
\put(1055,392){\raisebox{-.8pt}{\makebox(0,0){$\Box$}}}
\put(1072,392){\raisebox{-.8pt}{\makebox(0,0){$\Box$}}}
\put(1089,392){\raisebox{-.8pt}{\makebox(0,0){$\Box$}}}
\put(1106,392){\raisebox{-.8pt}{\makebox(0,0){$\Box$}}}
\put(1124,392){\raisebox{-.8pt}{\makebox(0,0){$\Box$}}}
\put(1141,392){\raisebox{-.8pt}{\makebox(0,0){$\Box$}}}
\put(1158,392){\raisebox{-.8pt}{\makebox(0,0){$\Box$}}}
\put(1175,392){\raisebox{-.8pt}{\makebox(0,0){$\Box$}}}
\put(1193,391){\raisebox{-.8pt}{\makebox(0,0){$\Box$}}}
\put(1210,391){\raisebox{-.8pt}{\makebox(0,0){$\Box$}}}
\put(1227,391){\raisebox{-.8pt}{\makebox(0,0){$\Box$}}}
\put(1244,391){\raisebox{-.8pt}{\makebox(0,0){$\Box$}}}
\put(1262,391){\raisebox{-.8pt}{\makebox(0,0){$\Box$}}}
\put(1279,391){\raisebox{-.8pt}{\makebox(0,0){$\Box$}}}
\put(1296,390){\raisebox{-.8pt}{\makebox(0,0){$\Box$}}}
\put(1313,390){\raisebox{-.8pt}{\makebox(0,0){$\Box$}}}
\end{Large}
\end{picture}
\vskip-0.6cm
\caption{Time history of the energy density in units of $\mu^4$
for $\mu=200{\rm MeV}$ (squares),
$\mu=100{\rm MeV}$ (pluses), and $\mu=50{\rm MeV}$ (diamonds).
Error bars are smaller than the plotting symbols. Proper time $\tau$ is
in fm.}
\label{ehistl160}
\end{figure}

We now turn to numerical results from our simulations. These were performed
for a variety of lattice sizes $L$= 20--160 
and the color charge density $\mu=0.025$--$0.3$ in 
units of the lattice spacing $a$. To convert lattice
results to physical units, we take $L^2 \approx \pi R^2$ for $A$=200 
nuclei. This then also 
determines $\mu$ for a fixed lattice size. The relation of lattice time to
continuum time is given by the relation 
$\tau_C = a\cdot \tau_L$~\cite{Krasnitz}. 
In Fig.~\ref{ekvsl}, we plot
the Gaussian averaged initial kinetic energy $\langle E_k\rangle$ 
on the lattice 
as a function of the lattice size $L$ and compare it 
with the lattice perturbation theory expression
\begin{eqnarray}
p^a p^a =&& N_c (N_c^2-1) \left(\mu\over {N}\right)^4 \\
&&\times\sum_{n,n^\prime} 
\left[ \left(\sum_l {\sin(l_n)\sin(l_{n^\prime})\over \Delta^2 (l)}
\right)^2 + 16\left(\sum_l {\sin^2({l_n\over 2})\sin^2({l_{n^\prime}
\over 2})\over \Delta^2 (l)}\right)^2 \right] \, , \nonumber
\label{dike}
\end{eqnarray}
where $\Delta(l) = 2 \sum_n[2 -\cos(2\pi l_n/L)]$, $n=x,y$, and 
$1-L/2\leq l_{x,y}\leq L/2$. The comparison is made
for values $\mu = 0.025, 0.05$ of the color charge density. For small
values of $L$, there is very good agreement between the two but for the
largest value $L=160$, they begin to deviate. The strong 
coupling parameter on the lattice 
is $\propto g^2\mu L$~\cite{RajGavai} and for $g^2 \mu L \gg 1$, 
we can expect to see deviations from lattice
perturbation theory.

In Fig.~\ref{a2kvspht025pil160mu025}, we plot the ratio of the field 
intensity of a particular
mode of the transverse gauge field as a function of proper time $\tau$
normalized to its value at $\tau=0$. The diamonds are results from a
lattice simulation with $L=160$ and $\mu =0.025$ and the mode
considered is $(k_x,k_y)=(\pi/4,0)$. Note that $k_{x,y}=2\pi
l_{x,y}/L$ and for this case, $l_x =20, l_y=0$. The solid line in the
figure is the square of the Bessel function $J_0 (\omega \tau)$ where
$\omega = \sqrt{\Delta(l)}$.
The time
dependence of the high transverse momentum modes should
agree with the continuum perturbative result 
which predicts a time dependence proportional to $J_0^2 (\omega\tau)$ 
for the field
intensity of the transverse gauge fields. The continuum dispersion relation
$\omega =|k_\perp|$ is however modified into the above mentioned 
lattice dispersion relation. 
We see from the figure that the anticipated agreement
between the lattice results and perturbation theory is quite good.

In Fig.~\ref{a2kvsphtminkl160}, we plot the same quantity as in Fig.~2, 
but now for three different 
values of $\mu$ and for the first non--zero momentum mode $(k_x,k_y) = (1,0)$.
The lattice size $L=160$ is the same as previously. For the smallest value of
$\mu = 0.025$, there is again an agreement with the Bessel behaviour predicted
by perturbation theory. However at the larger values of $\mu = 0.05, 0.1$, one
sees significant deviations away from the Bessel behaviour. Indeed the modes
appear to saturate at larger values of $\tau$. It is not clear that this 
saturation has much significance since the energy shows the
expected $1/\tau$ behaviour at late times. The late time behaviour 
of long wavelengths may be understood better by 
looking at gauge invariant quantities~\cite{RajKrasnitz2}.

In Fig.~\ref{na2vskl160}, the field intensity of the transverse gauge field 
normalized by $\mu^4$ at
$\tau =0$ is plotted as a function of the transverse momentum in 
physical units.  The lattice results for the different values of $\mu$ 
described in the caption
are compared to the lattice perturbation theory result below for the 
field intensity (the prime denotes that $\alpha$ satisfies the Coulomb 
gauge condition):
\begin{equation}
|\alpha_l^\prime|^2 = {{g^6 N_c (N_c^2-1) \mu^4}\over{4\Delta(l)}}\sum_{l'}
{{\Delta(2l'-l)\Delta(l)-\left[\Delta(l'-l)-\Delta(l')\right]^2}
\over{\Delta^2(l')\Delta^2(l'-l)}}\, .
\label{dcrfin}\end{equation}
The LPTh result (which would be the mini--jet distribution in the 
continuum) agrees very well with the lattice result for small $\mu$ up to
very small values of $k_t$. However, strong coupling effects grow with 
increasing $\mu$ (the lattice size $L$ is fixed here) and we see deviations
from the perturbative predictions at larger values of $k_t$. This trend is
enhanced further at larger values of $\mu$ than those shown here. The 
non--perturbative effects due to the non--linearities in the Yang--Mills 
equations seem to temper the $1/k_t^4$ behaviour predicted by perturbation 
theory. Whether this reflects the presence of a time--dependent 
mass in the theory needs further investigation.

Finally, we plot the time dependence of the energy density for different
values of $\mu$. At late times, from general considerations 
we expect that $E\propto 1/\tau$ and that is indeed what we see. 
Perturbation theory predicts the energy densities will scale as $\mu^4$.
Clearly our results do not show this scaling which suggests that
non--perturbative effects are important. It appears though that
the trend as we go to smaller values of $\mu$ is to approach the
perturbative scaling behaviour.

\section{Summary and outlook}

In this paper, we have described results from real time lattice simulations
of the full classical Yang--Mills equations with initial conditions given 
by the classical fields of each nucleus before the collision. At large 
transverse momenta, our simulations agree very well with lattice perturbation
theory as they should. At small transverse momenta, we show that there 
are significant deviations from the LPTh predictions. The late time behaviour
of these modes is particularly interesting. The spatial dependence of 
equal time correlators 
provide further insight into these modes. These will be discussed 
in a forthcoming paper~\cite{RajKrasnitz2}. While this work was being 
prepared, we received a preprint studying lattice simulations of nuclei in
a different approach~\cite{BMP}.
 
\section*{Acknowledgments}
We would like to thank the Universidade do Algarve (RV) and the
Niels Bohr Institute (AK) for
their kind hospitality. We would also like to thank Larry McLerran, 
Rob Pisarski and Dirk Rischke for very useful discussions. 
RV's work is supported by 
the Danish Research Council and the Niels Bohr Institute. Both AK and RV 
acknowledge support provided by the Portuguese Funda\c cao para a Ci\^encia e a 
Technologia,
grants CERN/S/FAE/1111/96 and CERN/P/FAE/1177/97.


\section*{References}
\begin{thebibliography}{99}
\bibitem{QM96}See, for instance, the proceedings of {\em Quark Matter 96},
\Journal{\NPA}{610}{}{1996}.

\bibitem{KajLanLin}K. Kajantie, P. V. Landshoff, and J. Lindfors, 
\Journal{\PRL}{59}{2527}{1987}; \\
K. J. Eskola, K. Kajantie, and J. Lindfors, \Journal{\NPB}{323}{37}{1989}; \\
J.-P. Blaizot and A. H. Mueller, \Journal{\NPB}{289}{847}{1987}.

\bibitem{Wang}X.-N. Wang, \Journal{\PRP}{280}{287}{1997}. 

\bibitem{Geiger}K. Geiger, \Journal{\PRP}{258}{237}{1995}.

\bibitem{GyuPluWang}M. Plumer, M. Gyulassy and X.-N. Wang, 
\Journal{\NPA}{590}{511c}{1995}.

\bibitem{KKV}K. J. Eskola, K. Kajantie, and P. V. Ruuskanen, 
\Journal{\EPJC}{1}{627}{1998}; \\
K. J. Eskola, \Journal{\em Comments in Nucl. and Part. Phys.}{22}{185}{1998}. 

\bibitem{comment}R. Venugopalan, 
\Journal{\em Comments in Nucl. and Part. Physics}{22}{113}{1998}.

\bibitem{GLR}L. V. Gribov, E. M. Levin, and M. G. Ryskin, 
\Journal{\em Phys. Rep.}{100}{1}{1983}.

\bibitem{RajLar}
L. McLerran and  R. Venugopalan, \Journal{PRD}{49}{2233}{1994}; \\
\Journal{\PRD}{49}{3352}{1994}; \\
\Journal{\PRD}{50}{2225}{1994}.

\bibitem{AJMV}A. Ayala, J. Jalilian-Marian, L. McLerran and R. Venugopalan,
\Journal{\PRD}{52}{2935}{1995}; \\
\Journal{\PRD}{53}{458}{1996}.

\bibitem{JKMW}J. Jalilian-Marian, A. Kovner, L. McLerran and H. Weigert, 
\Journal{\PRD}{55}{5414}{1997}.

\bibitem{JKLW}J. Jalilian-Marian, A. Kovner, A. Leonidov and H. Weigert,
\Journal{\NPB}{504}{415}{1997}; \\
hep-ph/9807462.

\bibitem{JKW}J. Jalilian-Marian, A. Kovner, and H. Weigert, hep-ph/9709432.

\bibitem{Kovchegov}Yu. V. Kovchegov, 
\Journal{\PRD}{54}{5463}{1996}; \\
\Journal{\PRD}{55}{5445}{1997}.

\bibitem{KLW}A. Kovner, L. McLerran and H. Weigert, 
\Journal{\PRD}{52}{3809}{1995}; \\
\Journal{\PRD}{52}{6231}{1995}.

\bibitem{gyulassy} M. Gyulassy and L. McLerran, 
\Journal{{\em Phys. Rev.} C}{56}{2219}{1997}.

\bibitem{DirkYuri}Y. V. Kovchegov and D. H. Rischke, 
\Journal{{\em Phys. Rev.} C}{56}{1084}{1997}.

\bibitem{SerBerDir}S. G. Matinyan, B. M\"uller and D. H. Rischke, 
\Journal{{\em Phys. Rev.} C}{56}{2191}{1997}.

\bibitem{Bj}J. D. Bjorken, \Journal{\PRD}{27}{140}{1983}; \\
J. Sollfrank, P. Houvinen, M. Kataja, P. V. Ruuskanen, 
M. Prakash and R. Venugopalan, 
\Journal{{\em Phys. Rev.} C}{55}{392}{1997}.

\bibitem{GunionBertsch}J. F. Gunion and G. Bertsch, 
\Journal{\PRD}{25}{746}{1982}.

\bibitem{RajKrasnitz}A. Krasnitz and R. Venugopalan, hep-ph/9706329, 
in proceedings of {\it 3rd International Conference on the Physics and
Astrophysics of the Quark Gluon Plasma}, March 17th--21st, Jaipur, India. 

\bibitem{RajKrasnitz2}A. Krasnitz and R. Venugopalan, NBI-98-21,
UALG/PT/5, to be published.

\bibitem{Krasnitz}J. Ambj{\o}rn and A. Krasnitz, 
\Journal{\PLB}{362}{97}{1995}; \\
\Journal{\NPB}{506}{387}{1997}.

\bibitem{Moore}G.D. Moore, 
\Journal{\NPB}{480}{657}{1996}.

\bibitem{MuellKovWall}A.H. Mueller, Y. Kovchegov, and S. Wallon, 
\Journal{\NPB}{507}{367}{1997}; \\
A. H. Mueller and Y. Kovchegov, hep-ph/9802440.

\bibitem{Mueller}A.H. Mueller, 
\Journal{\NPB}{415}{373}{1994}.

\bibitem{MakhSurd}A. Makhlin and E. Surdutovich, 
\Journal{{\em Phys. Rev.} C}{58}{389}{1998}.

\bibitem{Ian}I. Balitsky, hep-ph/9808215.

\bibitem{EskMullWang}K. J. Eskola, B. M\"{u}ller, and X.-N. Wang, 
\Journal{\PLB}{374}{20}{1996}.

\bibitem{GyuWang}M. Gyulassy and X.-N. Wang, 
\Journal{\NPB}{420}{583}{1994}.

\bibitem{Sasha}A. Makhlin, hep-ph/9608261.

\bibitem{KS}J. Kogut and L. Susskind, 
\Journal{\PRD}{11}{395}{1975}.

\bibitem{RajGavai}R. V. Gavai and R. Venugopalan, 
\Journal{\PRD}{54}{5795}{1996}.

\bibitem{BMP}S. A. Bass, B. M\"uller and W. P\"oschl, nucl-th/9808011.

\end{thebibliography}
\end{document}